\documentclass[]{scrartcl}

\usepackage{bm}
\usepackage{xcolor}
\usepackage{comment}
\usepackage{hyperref}
\usepackage{listings}
\usepackage{colortbl}
\usepackage{listings}
\usepackage[utf8]{inputenc}
\usepackage{amssymb,amsmath,amsthm}

\theoremstyle{definition}
\newtheorem{example}{Example}[section]

\DeclareFixedFont{\ttb}{T1}{txtt}{bx}{n}{10} 
\DeclareFixedFont{\ttm}{T1}{txtt}{m}{n}{10}  

\definecolor{headcolor}{cmyk}{0.77,0.27,0.,0.13}
\definecolor{rowcolor}{cmyk}{0.4,0.1,0.0,0.0}

\definecolor{deepblue}{rgb}{0,0,0.5}
\definecolor{deepred}{rgb}{0.6,0,0}
\definecolor{deepgreen}{rgb}{0,0.5,0}
\definecolor{deeporange}{rgb}{255,140,0}

\newcommand\cheasepystyle{\lstset{
		language=Python,
		basicstyle=\ttm,
		otherkeywords={setParam,namelistVals,**kwargs,profilesfpath,
		               exptnzfpath,expeqfpath,eqdskfpath,iterdbfpath,
	                   cheasefpath,Normalized,srcVals,cheaseVals,outfile,
                       importedVals,namelistVals, boundary_type,psi,eps,
                       array,runmode,cheasemode,removeinputs,removeoutputs,shotpath},
		keywordstyle=\ttb\color{deepblue},
		emph={create_namelist,read_profiles,read_exptnz,
		      read_expeq,read_eqdsk,read_iterdb,read_chease,
	          write_exptnz,write_expeq,read_efit_file,python,
              cheasepy,magsurf_solvflines,plot_chease},
		emphstyle=\ttb\color{deepred},
		stringstyle=\ttb\color{deepgreen},
		showstringspaces=false
}}
\lstnewenvironment{cheasepy}[1][]
{\cheasepystyle\lstset{#1}}{}

\newcommand\cheasepyinline[1]{{\cheasepystyle\lstinline!#1!}}

\title{CheasePy\\ \Large User Guide}
\author{Ehab Hassan}

\begin{document}

\maketitle

\begin{abstract}
CheasePy is code written in Python to run the CHEASE (\textit{Cubic Hermite Element Axisymmetric Static Equilibrium}) code, which solves the Grad-Shafranov equation for toroidal MHD equilibria using pressure and current profiles and fixed plasma boundaries that is defined by a set of experimental data points (R,Z). The CheasePy code allows an iterative running of the CHEASE code either to check the preservation of MHD equilibria or converging to an experimentally defined total toroidal plasma current by modifying any input quantity.
\end{abstract}

\section{Toroidal MHD Equilibrium}
\subsection{Grad-Shafranov Equation}\label{GradShafEq}
The MHD equilibrium equations are given by:
\begin{align}
\bm{J\times B}&=\bm{\nabla}p \\
\bm{\nabla\times B}&=\bm{J} \\
\bm{\nabla\cdot B}&=0
\end{align}
The magnetic field can be represented as:
\begin{equation}
\textbf{B}=T\bm{\nabla}+\bm{\nabla}\phi\times\bm{\nabla}\Psi
\end{equation}
where $\phi$ is the ignorable toroidal angle and $\Psi$ is the poloidal magnetic flux function.\\~\\
For static MHD equilibria, the Grad-Shafranov equation is given by:
\begin{equation}
\bm{\nabla\cdot}\frac{1}{R^2}\bm{\nabla}\Psi = \frac{j_{\phi}}{R} = -p'(\Psi)-\frac{1}{R^2}TT'(\Psi)
\end{equation}
where $j_{\phi}$ denotes the toroidal plasma current density, R the major radius of the torus. The nature of the equilibria is determined by the two free functions $p'(\Psi)$ and $TT'(\Psi)$, where the pressure p and the poloidal current flux function T are functions of $\Psi$ only.\\~\\
The total current everywhere inside the plasma ($\Psi\textless 0$) is positive and is given by:
\begin{equation}
I=\int{j_{\phi} dS} = \int{j_{\phi} \left(J/R\right) d\Psi d\chi}
\end{equation}
where J is the Jacobian.\\~\\
\newpage
\subsection{Current Profiles}\label{CurrentProf}
To achieve MHD equilibrium CHEASE can treat different current profiles such as:
\begin{itemize}
	\item The poloidal current flux:
	\begin{equation}
	TT'(\Psi)
	\end{equation}
	\item The surface averaged current density:
	\begin{equation}
	I^*(\Psi)=\frac{\oint j_{\phi}(J/R)d\chi}{\oint (J/R)d\chi}=-R_0^2\frac{C_1}{C_0}p'(\Psi)-R_0^2\frac{C_2}{C_0}\frac{TT'(\Psi)}{\mu_0}
	\end{equation}
	\item The averaged parallel current:
	\begin{align}
	I_{\parallel}&=\frac{\oint \bm{J\cdot B} Jd\chi}{\oint \bm{B\cdot\nabla}\phi Jd\chi}=R_0\frac{\left<\bm{J\cdot B}\right>}{\left<T/R^2\right>} \label{right} \\
	I_{\parallel}&=\frac{\oint \bm{J\cdot B} Jd\chi}{\oint \bm{B\cdot\nabla}\phi Jd\chi}=-R_0\frac{C_1}{C_2}p'(\Psi)-R_0\frac{TT'(\Psi)}{\mu_0}(\Psi)\left(1+\frac{1}{T^2(\Psi)}\frac{C_3}{C_2}\right) \label{wrong}
	\end{align}
	\item The averaged parallel current density:
	\begin{equation}
	J_{\parallel}=\frac{\left<\bm{J\cdot B}\right>}{B_0}
	\end{equation} 
\end{itemize}
where,
\begin{align}
\left\{C_0,C_1,C_2,C_3\right\}&=\oint \left\{\frac{1}{R},1,\frac{1}{R^2},\frac{|\nabla\Psi|^2}{R^2}\right\}Jd\chi \\
\left<\bm{J\cdot B}\right>&=-T(\Psi)p'(\Psi)-T'(\Psi)\frac{\left<B^2\right>}{\mu_0} \\
\left<B^2\right>&=\frac{\oint B^2 Jd\chi}{\oint Jd\chi}\\
\left<T/R^2\right>&=\frac{\oint T/R^2 Jd\chi}{\oint Jd\chi}
\end{align}
The toroidal current density can be expressed as:
\begin{align}
j_{\phi}&=\frac{1}{R} \left(\frac{C_0}{C_2}I^*(\Psi)+\left(\frac{C_1}{C_2}-R^2\right)p'(\Psi)\right) \\
j_{\phi}&=\frac{1}{yR}\left(I_{\parallel}(\Psi)+\left(\frac{C_1}{C_2}-yR^2\right)p'(\Psi)\right)
\end{align}
where \label{wrong}
\begin{equation}
y = 1+\frac{1}{T^2(\Psi)}\frac{C_3}{C_2}
\end{equation}
\subsection{Normalized Quantities}\label{Normalize}
The normalization used here has the following format:
\begin{align}
\Psi&=B_0R_0^2 \nonumber\\
I&=\frac{B_0R_0}{\mu_0} ~~~~~ and ~~~~~ J=\frac{B_0}{\mu_0R_0} \nonumber\\
p&=\frac{B_0^2}{\mu_0} ~~~~~~~~ and ~~~~~ p'=\frac{B_0}{\mu_0R_0^2} \nonumber\\
T&=B_0R_0 ~~~~~ and ~~~~~ T'=\frac{1}{R_0} ~~~~~~~~ and ~~~~~ TT'=B_0 \nonumber
\end{align}

\newpage
\section{CHEASE [Fortran] Code}
When CHEASE code runs based on experimental data points it takes as an input three files. The \textit{EXPEQ} file (which is a kind of EFIT file, also known as EQDSK file) which has the experimental equilibrium quantities such as the current profile, in addition to the magnetic configuration. The second file is the \textit{EXPTNZ} file which contains profiles for the ion and electron temperatures ($T_i~\&~T_e$) and densities ($n_i~\&~n_e$), in addition to the effective atomic number ($Z_{eff}$) as a function of the normalized poloidal axis ($\psi_N$). The CHEASE code also takes the \textit{chease\_namelist} file which has the initialization parameters that depend on the plasma fusion machine shot, the type of the input physical quantities, and the run-mode of the CHEASE code. For more information about CHEASE code and the different options for the input parameters in the \textit{chease\_namelist} file you may go to their \textcolor{blue}{\href{https://crppwww.epfl.ch/~sauter/chease/}{website}} which has all the required resources and references. \\~\\
After a successful run, the CHEASE code produces several output files, but CheasePy code uses only \textit{four} of them. The \textit{EXPEQ.OUT(IN)} output files have specific experimental quantities from the EFIT file depending on the selected values of the input parameters in the \textit{chease\_namelist} file, such as pressure ($P$), poloidal coordinate grid points ($\rho(\psi)$), parallel current density profile ($J_{\parallel}$), etc. \textit{EXPTNZ.OUT(IN)} output file should have exactly the same profiles in the \textit{EXPTNZ} input file but with a higher resolution. CHEASE code also packs several output quantities related to the coordinate systems and physical quantities into an HDF5 file called \textit{ogyropsi.h5}, and it should be noticed here that all the quantities in the HDF5 file have SI units in contrary to the normalized quantities in the \textit{EXPEQ.OUT(IN)} files. CHEASE code also gives a text file as an output that contains all the outcomes similar to the HDF5 file named based on the iteration number to be \textit{ogyropsi.dat}. The CHEASE code produces also several other output files but they aren't needed by CHEASEPY.\footnote{CheasePy renames all the required input and output files to include the iteration number at the end of each filename, e.g. CheasePy renames EXPEQ.OUT to EXPEQ\_iter000.OUT after the zeroth iteration and does the same after each iteration for all IN and OUT files. However, in case of the files ogyropsi.dat and ogyropsi.h5, CheasePy renames them to chease\_iterxxx.dat and chease\_iterxxx.h5, respectively, after each iteration.}\\~\\
It should be noticed here that for calculating the MHD equilibrium in the zero iteration the CHEASE code uses an EXPEQ file that may have the EFIT or EQDSK format which contains an extensive amount of information about the machine geometry, magnetic configuration, and plasma status. However it can also use a reduced format of the EQDSK file which contains only the required data for the next iteration to run properly. In contrary, the format of the EXPTNZ doesn't change from one iteration to another, it has the same profiles but with higher resolution as mentioned above. Some parameters in the \textit{chease\_namelist} file need to be changed from one iteration to another, but most of them keep their values. For example, the \textit{NEQDSK} parameter takes 1 when using the EFIT file in EQDSK format, or it takes 0 to create the EXPEQ from input/given parameters. We will shed more lights on the description and function of the parameters needed to be modified to properly run the CheasePy code in other sections, but it worth mentioning here that the provided \textit{chease\_namelist} file with the CheasePy code has only the parameters that you may need to change based on the case you run such as pressure type (nppfun), current type (nsttp), coordinates resolution (NPSI,NCHI,NS,NT,NISO). Other parameters which are not provided in the basic \textit{chease\_namelist} file are set to their default values by CheasePy code in the \textit{namelistcreate()} function, or by the CHEASE code itself. You may also add these parameters to the \textit{chease\_namelist} file in case you will change them in a regular basis.

\newpage
\section{CheasePy [Python] Code}
To run the CheasePy code you need to have the CHEASE code compiled first in your local machine as instructed in the CHEASE developer \textcolor{blue}{\href{https://crppwww.epfl.ch/~sauter/chease/}{website}} and have the executable file named \textit{chease\_hdf5} in the same directory\footnote{It is recommended to set/export the \$PATH to the location of the \textit{cheasepy.py} and \textit{efittools.py} files, and to keep the \textit{runchease.py} file with the \textit{chease\_hdf5} executable in the same folder.} of the \textit{cheasepy.py}\footnote{cheasepy.py is a Python package that contains all the required functions to perform all operations in the CheasePy code such as profile and geometry files creation, profile and geoemtry files reading, etc.}, \textit{runcchease.py}\footnote{runchease.py is a Python script that allows the user of CheasePy to setup the sources and types of all input profiles and geometry, and it also allows the user to setup the mode of operation of CheasePy code.}, and \textit{efittools.py}\footnote{efittools.py is a Python package that contains functions to read the efit, profile, and iterdb files, and calculate the outer most closed magnetic surface that does not have the x-point.} files. Also, you need to have \textcolor{blue}{\href{https://www.anaconda.com/download/}{Anaconda for Python 2.7/3.7 version}} installed in your local machine, where the \textit{cheasepy.py} script works with both versions of Python (2.7 or 3.7).
\subsection{Directory Structure}
In the main directory there should be at least two files, \textit{runchease.py} and \textit{chease\_hdf5} with the path to \textit{cheasepy.py} and \textit{efittools.py} files being accessible. Also, the path to the files of the desired shot can be set to the `\textit{shotpath}' variable in the \textit{runchease.py} script.\\~\\
The \textit{\textbf{shots}} folder contains subfolders for each available experimental output measurements. Each shot subfolder is named upon the user choice, but it is preferred to be named after the machine/shot name, e.g. \textit{machine.name\_shot.number}, and contains at least a profile file and a equilibrium geometry file with their names formatted to have `\textit{the parent folder name}' then `\textit{underscore}' then the type of data file, e.g. \textit{machine.name\_shot.number\_FILE.TYPE}:
\begin{center}
  \small
  \begin{tabular}{|c|l|}
  	\hline
  	\rowcolor{headcolor}
  	FLE.TYPE & Description \\
  	\hline
  	\rowcolor{rowcolor}
	EQDSK & Geometry Equilibrium and Current and Pressure Profiles (full) \\
	\rowcolor{rowcolor}
	EXPEQ & Geometry Equilibrium and Current and Pressure Profiles (reduced) \\
	\rowcolor{rowcolor}
	EXPTNZ & Electron and Ions Profiles (reduced) \\
	\rowcolor{rowcolor}
	ITERDB & Electron and Ions Profiles (reduced) \\
	\rowcolor{rowcolor}
	PROFILES & A Complete Equilibrium Profiles for Ions, Electrons, Impurities, etc. (full) \\
	\rowcolor{rowcolor}
	CHEASE & A Complete Geometry and Profiles output from a previous Chease run (full) \\
	\hline
  \end{tabular}
\end{center}
You may add more shots to the shot folder by following the same method of naming the shot folder and the equilibrium geometry and profile files. Having a wrong name for the shot folder and/or any of its contents might cause \textit{CheasePy} code doesn't work properly.\\~\\
CheasePy has the capability to take the electron profiles and ion profiles from different sources, it can also take the pressure profile from the EQDSK (EFIT) file or calculate it from any of the input profile sources.\footnote{Calculating the pressure from EXPTNZ or ITERDB ignores the contribution of the fast particles ($n_b$).} Because the profiles on each source are projected on its own grid, CheasePy allows users to unify all the provided profiles to a single grid by setting a source for the rhomesh grid.\footnote{Considering both $\rho_{\psi_N}$ and $\rho_{\phi_N}$ coordinates, CheasePy prefers a rhomesh grid source that has both coordinates to ease the interpolation process in different sources.}

\newpage
\subsection{Creating {chease\_namelist} File}\label{CreateNamelist}
The \textit{CheasePy} code starts with copying the required profiles and equilibrium geometry files of the selected shot to the current directory, and then it uses the user-defined namelist parameters in \textit{runchease.py} script file to create the \textit{chease\_namelist} file (which is required for CHEASE code to run as we indicated above) using the \cheasepyinline{create\_namelist(setParam=\{\})} function in the \textit{cheasepy.py} package.\\~\\
There is a long list of CHEASE namelist parameters that are used to set the operation mode of CHEASE code, but working on experimental measurements helps in reducing this list to smaller number of parameters, with all other parameters to set to their default values defined by either CHEASE or CheasePy. A list of the required CHEASE parameters can be found in the \textit{runchease.py} script, and they can be modified based on the provided experimental facts. The namelist parameters are stored in a Python dictionary structure called \cheasepyinline{namelistVals}\footnote{The \cheasepyinline{namelistVals} is passed to \cheasepyinline{setParam} argument in \cheasepyinline{create\_namelist} function.} with the namelist parameters are the dictionary keys named as defined in CHEASE code. The value of each namelist parameter can be \textit{single-valued} or \textit{list-of-values} depending on the number of times the input namelist parameters will change from one iteration of running CHEASE to another. For example, if CheasePy runs CHEASE for 11 times but each parameter to the \cheasepyinline{namelistVals} is \textit{single-valued}, then the same namelist input file created in the zeroth iteration will be used in the other 10 iterations. However, if the parameters in the \cheasepyinline{namelistVals} are defined as a list of two values, the first value will be used for the zeroth iteration of CHEASE then the second value will be used for the other 10 iterations. Generally, \textit{CheasePy} will create a new \textit{chease\_namelist} file based on the number of values in the \cheasepyinline{namelistVals} and use the latest one for the rest of iterations.\\~\\
The description of the CHEASE namelist parameters available in the \textit{runchease.py} script can be found in the following table:\footnote{For a full list of the CHEASE input parameters and their detailed description you may read this reference: \textcolor{blue}{\href{https://www.sciencedirect.com/science/article/pii/001046559600046X}{The CHEASE code for toroidal MHD equilibria}} by H. Lutjens, A. Bondeson, O. Sauster} \footnote{You may read more details about how to choose the values of \textit{cocos\_in} and \textit{cocos\_out} in the \textcolor{blue}{\href{https://www.sciencedirect.com/science/article/pii/S0010465512002962}{Tokamak coordinate conventions: COCOS}} by O.Sauter and S.Yu.Medvedevb.}
\begin{center}
\small
\begin{tabular}{|c | c | l |}
\hline
\rowcolor{headcolor}
Parameter & Default & Description \\
\hline
\rowcolor{rowcolor}
NS & 256 & Number of tr-intervals for equilibrium calculation. \\
\rowcolor{rowcolor}
NT & 256 & Number of 0-intervals for equilibrium calculation. \\
\rowcolor{rowcolor}
NISO & 256 & Number of s-intervals to define $I^{*}(s)$ or $I_{||}(s)$. \\
\rowcolor{rowcolor}
NPSI & 1024 & Number of radial stability-s intervals.\\
\rowcolor{rowcolor}
NCHI & 1024 & Number of poloidal nodes for ballooning. \\
\rowcolor{rowcolor}
NRBOX & 60 & Number of R points used to save equilibrium in EQDSK. \\
\rowcolor{rowcolor}
NZBOX & 60 & Number of Z points used to save equilibrium in EQDSK. \\
\rowcolor{rowcolor}
RELAX & 0 & Under-relaxation parameter used if magnetic axis converges slowly \\
\rowcolor{rowcolor}
NSTTP & 1 & Type of input current profiles: 1=$ff^{'}$, 2=$I^{*}$, 3=$I_{||}$, and 4=$J_{||}$. \\
\rowcolor{rowcolor}
NPROPT & 3 & Type of output current profiles: 1=$ff^{'}$, 2=$I^{*}$, 3=$I_{||}$, and 4=$J_{||}$. \\
\rowcolor{rowcolor}
NPPFUN & 8 & Type of input pressure profiles: 4=$P^{'}$ and 8=$P$. \\
\rowcolor{rowcolor}
NEQDSK & 0 & Source of equilibrium geometry: 0=EXPEQ and 1=EQDSK \\
\rowcolor{rowcolor}
TENSBND & 0 & Interpolation weight around the x-point. \\
\rowcolor{rowcolor}
COCOS\_IN & 2 & To determine the poloidal coordinate systems from input EXPEQ. \\
\rowcolor{rowcolor}
TENSPROF & 0 & Interpolation weight of the profiles. \\
\rowcolor{rowcolor}
COCOS\_OUT & 12 & To determine the poloidal coordinate systems from output EXPEQ. \\
\rowcolor{rowcolor}
NRHOMESH & 0 & Type of input grid: 0=$\rho_{\psi_N}$ and 1=$\rho_{\phi_N}$.\\
\hline
\end{tabular}
\end{center}
The \textit{TENSBND} parameter is set to a negative value, e.g. -1, when running CHEASE with NEQDSK=1, i.e. using the EQDSK file as a source for the equilibrium geometry such as the (R,Z) coordinates for the magnetic boundary surface to eliminate the x-point by interpolation. This is not applicable when with NEQDSK=0 because a EXPEQ file will be constructed in a different way that is explained below.

\newpage
\subsection{Reading Input/Output Files}
There are different input/output files that can be used with \textit{CheasePy} such as EQDSK, EXPEQ, EXPTNZ, PROFILES, ITERDB, and CHEASE. Each of these files organizes the data/measurements in a different way, and each input/output file has its own grid type and scale. In the following subsection we will show how a user can use \textit{CheasePy} to read these input/output files, and get \textit{CheasePy} to interpolate these read data/measurements on another grid specified/provided by the user.
\subsubsection{Reading `Profiles' File}\label{readProfiles}
The equilibrium profiles of different quantities measured in the fusion reactors, such as electron, ion, impurities, and fast ion densities, toroidal and poloidal speeds, etc. are provided by the sentimentalists, and are usually projected on a $\psi_N$ normalized grid and have the same equilibrium for the physical and geometrical quantities in the \textit{EFIT} file.\\~\\
\textit{CheasePy} provides the \cheasepyinline{read\_profiles(profilesfpath,setParam,**kwargs)} function to read the \textit{profiles} file. This function takes the following arguments:
\begin{center}
	\small
	\begin{tabular}{|c | c | l |}
		\hline
		\rowcolor{headcolor}
		Argument & Default & Description \\
		\hline
		\rowcolor{rowcolor}
		exptnzfpath & user-input & The path to the \textit{profile} file. \\
		\rowcolor{rowcolor}
		setParam & \{\} & nrhopsi=0 ($\rho_{\psi_N}$) and nrhopsi=1 ($\rho_{\phi_N}$) grid. \\
		\rowcolor{rowcolor}
		~ & ~ & Zeff=True for $Z_{eff}$ profile or Zeff=False for fixed-value $Z_{eff}$. \\
		\rowcolor{rowcolor}
		**kwargs & None & Choose a source for the $\rho_{\psi_N}$ and $\rho_{\phi_N}$ grid to interpolate on.\\
		\hline
	\end{tabular}
\end{center}
The first argument, \cheasepyinline{profilesfpath}, takes the path to the \textit{profiles} file which \textbf{must be} provided by the user. If the second argument, \cheasepyinline{setParam}, is an empty dictionary, the \cheasepyinline{read\_profiles} function returns all quantities projected on the original grid ($\psi_N$) provided in the profiles file with the original $Z_{eff}$ profile calculated using this equation:
\begin{equation}
     Z_{eff} = \frac{\sum_{s}n_sZ_s^2}{n_e}
\end{equation}
However, if \textbf{Zeff} key is set in \cheasepyinline{setParam} to False, all grid points of the $Z_{eff}$ profile is set to the average value of the effective atomic number profile calculated using the equation above.\\~\\
A source for $\psi_N$ ($\rho_{\psi_N}$) and $\phi_N$ ($\rho_{\phi_N}$) should be provided in the \cheasepyinline{**kwargs} input argument, hence setting \textbf{nrhopsi} in \cheasepyinline{setParam} to 0 will return all the profiles interpolated into the new $\psi_N$ ($\rho_{\psi_N}$) grid provided in the source file, however all the profiles will be interpolated into the new $\phi_N$ counterpart if \textbf{nrhopsi} is set to 1, and \cheasepyinline{read\_profiles} function will return only the new $\rho_{\psi_N}$, $\rho_{\phi_N}$, $\psi_N$, and $\phi_N$ grids to the calling function/script.\\~\\
To provide a source for the grid points, we add the type of that source and a path to the source file to the \cheasepyinline{**kwargs} input in the \cheasepyinline{read\_profiles} function call. Because the user may request to interpolate the profiles onto $\psi_N$ ($\rho_{\psi_N}$) and $\phi_N$ ($\rho_{\phi_N}$) grids the source must provide both grids point, otherwise \textit{CheasePy} will \cheasepyinline{raise} an error in case the interpolation onto the $\phi_N$ ($\rho_{\phi_N}$) grid is requested but the input source doesn't provide both grid points to implement that interpolation.\footnote{If the user wants to interpolate the profiles onto a new grid of the same type, the source may have only the new grid points, e.g. $\psi_N$ ($\rho_{\psi_N}$), and provided it to the \cheasepyinline{**kwargs} argument in \cheasepyinline{read\_profiles} function.} Therefore, the type of the sources allowed for interpolating onto a grid of different type, e.g. $\phi_N$ ($\rho_{\phi_N}$), are limited to \textit{CHEASE}, \textit{EQDSK}, and \textit{IMPORTED}.\footnote{Limiting the sources for the grid to these sources is because they have the q-profile which make it possible to convert from $\psi_N$ ($\rho_{\psi_N}$) to $\phi_N$ ($\rho_{\phi_N}$) using this formula: $\phi = \int \psi dq$.} The last source allows you to import $\psi_N$ ($\rho_{\psi_N}$) and $\phi_N$ ($\rho_{\phi_N}$) directly from any external source, and they will override any other grid provided in \textit{CheasePy}.\footnote{The \textbf{imported} values is explained more in the description of running \textit{CheasePy} and setting its mode of operation in the \textit{runchease.py} file.}\\~\\
The \cheasepyinline{read\_profiles} function returns all the profiles encapsulated in a dictionary in SI units. The name (=key) and description of the returned profiles are listed in the following table:
\begin{center}
	\small
	\begin{tabular}{|c | l |}
		\hline
		\rowcolor{headcolor}
		Measurement & Description \\
		\hline
		\rowcolor{rowcolor}
		Te & Electron Temperature \\
		\rowcolor{rowcolor}
		Ti & Ion Temperature \\
		\rowcolor{rowcolor}
		Tb & Fast Ions Temperature \\
		\rowcolor{rowcolor}
		ne & Electron Density \\
		\rowcolor{rowcolor}
		ni & Ion Density \\
		\rowcolor{rowcolor}
		nb & Fast Ions Density \\
		\rowcolor{rowcolor}
		nz & Impurities Density \\
		\rowcolor{rowcolor}
		Pb & Fast Ions Pressure \\
		\rowcolor{rowcolor}
		Zeff & Effective Atomic Number \\
		\rowcolor{rowcolor}
		Vtor & Toroidal Velocity \\
		\rowcolor{rowcolor}
		Vpol & Poloidal Velocity \\
		\rowcolor{rowcolor}
		PSIN & $\psi_N$ \\
		\rowcolor{rowcolor}
		PHIN & $\phi_N$ (with grid source) \\
		\rowcolor{rowcolor}
		rhopsi & $\rho_{\psi_N}$ \\
		\rowcolor{rowcolor}
		rhotor & $\rho_{\phi_N}$ (with grid source) \\
		\rowcolor{rowcolor}
		pprime & Total Pressure Gradient \\
		\rowcolor{rowcolor}
		pressure & Total Pressure of All Species \\
		\hline
	\end{tabular}
\end{center}

\subsubsection{Reading `EXPTNZ' File}\label{readEXPTNZ}
The \textit{EXPTNZ} file is an INPUT/OUTPUT file for \textit{CHEASE} code to provide the density and temperature profiles of the ions and electrons in SI units. The first line of the EXPTNZ file shows the number of grid points, grid type, i.e. rhopsi ($\rho_{\psi_N}$) or rhotor ($\rho_{\phi_N}$), and the order of the given profiles, such as electron temperature ($T_e$), electron density ($n_e$), effective atomic number ($Z_{eff}$), ion temperature ($T_e$), and ion density ($n_i$), in the file. These profiles are organized in a single column with each of them has a length equals the given number of grid points. The header of a \textit{EXPTNZ} file looks like:
\begin{center}
	\texttt{512 rhopsi,  Te,   ne,   Zeff,   Ti,   ni  profiles}
\end{center}
\textit{CheasePy} provides the \cheasepyinline{read\_exptnz(exptnzfpath,setParam,**kwargs)} function to read the \textit{EXPTNZ} file.This function takes the following arguments:
\begin{center}
	\small
	\begin{tabular}{|c | c | l |}
		\hline
		\rowcolor{headcolor}
		Argument & Default & Description \\
		\hline
		\rowcolor{rowcolor}
		exptnzfpath & user-input & The path to the \textit{EXPTNZ} file. \\
		\rowcolor{rowcolor}
		setParam & \{\} & Select between nrhopsi=0 ($\rho_{\psi_N}$) and nrhopsi=1 ($\rho_{\phi_N}$) grid. \\
		\rowcolor{rowcolor}
		**kwargs & None & Choose a source for the $\rho_{\psi_N}$ and $\rho_{\phi_N}$ grid to interpolate on.\\
		\hline
	\end{tabular}
\end{center}
The first argument, \cheasepyinline{exptnzfpath}, takes the path to the \textit{EXPTNZ} file which \textbf{must be} provided by the user. If the second argument, \cheasepyinline{setParam}, is an empty dictionary, the \cheasepyinline{read\_exptnz} function returns all quantities projected on the original grid ($\rho_{\psi_N}$) provided in the \textit{EXPTNZ} file, otherwise, a source for $\rho_{\psi_N}$ and $\rho_{\phi_N}$ should be provided in the \cheasepyinline{**kwargs} input argument, hence setting \textbf{nrhopsi} in \cheasepyinline{setParam} to 0 will return all the profiles interpolated into the new $\rho_{\psi_N}$ grid provided in the source file, however all the profiles will be interpolated into the new $\rho_{\phi_N}$ counterpart if \textbf{nrhopsi} is set to 1, and \cheasepyinline{read\_exptnz} function will return only the new $\rho_{\psi_N}$ and $\rho_{\phi_N}$ grids to the calling function/script and ignore the original grids provided in the \textit{EXPTNZ} file. The source of new grids are passed to the \cheasepyinline{**kwargs} input argument of the \cheasepyinline{read\_exptnz} function as described in section(\ref{readProfiles}).\\~\\
In addition to the input profiles, \cheasepyinline{read\_exptnz} function calculates and returns the density and temperature of the impurities and the total pressure. The total pressure calculated in the \cheasepyinline{read\_exptnz} doesn't include the contribution of the fast ions (if any), which might make the pressure calculated from the \textit{EXPTNZ} file is different from the actual pressure measured in the experiment.\footnote{The actual total pressure measured in the experiment is found in the \textit{profiles} and \textit{EQDSK} files.}\\~\\
The \cheasepyinline{read\_exptnz} function returns all the profiles encapsulated in a dictionary in SI units. The name (=key) and description of the returned profiles are listed in the following table:
\begin{center}
	\small
	\begin{tabular}{|c | l |}
		\hline
		\rowcolor{headcolor}
		Measurement & Description \\
		\hline
		\rowcolor{rowcolor}
		Te & Electron Temperature \\
		\rowcolor{rowcolor}
		Ti & Ion Temperature \\
		\rowcolor{rowcolor}
		ne & Electron Density \\
		\rowcolor{rowcolor}
		ni & Ion Density \\
		\rowcolor{rowcolor}
		nz & Impurities Density \\
		\rowcolor{rowcolor}
		Zeff & Effective Atomic Number \\
		\rowcolor{rowcolor}
		rhopsi & $\rho_{\psi_N}$ \\
		\rowcolor{rowcolor}
		rhotor & $\rho_{\phi_N}$ (with grid source) \\
		\rowcolor{rowcolor}
		pprime & Total Pressure Gradient \\
		\rowcolor{rowcolor}
		pressure & Total Pressure of All Species \\
		\hline
	\end{tabular}
\end{center}

\subsubsection{Reading `ITERDB' File}\label{readITERDB}
The \textit{ITERDB} file provides the density and temperature profiles for electrons, ions, and impurities, in addition to the toroidal velocity defined on a \textit{rhotor} (i.e. $\rho_{\phi_N}$) grid and SI units.\\~\\
\textit{CheasePy} provides the \cheasepyinline{read\_iterdb(iterdbfpath,setParam,**kwargs)} function to read the \textit{ITERDB} file.This function takes the following arguments:
\begin{center}
	\small
	\begin{tabular}{|c | c | l |}
		\hline
		\rowcolor{headcolor}
		Argument & Default & Description \\
		\hline
		\rowcolor{rowcolor}
		iterdbfpath & user-input & The path to the \textit{ITERDB} file. \\
		\rowcolor{rowcolor}
		setParam & \{\} & Select between nrhopsi=0 ($\rho_{\psi_N}$) and nrhopsi=1 ($\rho_{\phi_N}$) grid. \\
		\rowcolor{rowcolor}
		**kwargs & None & Choose a source for the $\rho_{\psi_N}$ and $\rho_{\phi_N}$ grid to interpolate on.\\
		\hline
	\end{tabular}
\end{center}
The first argument, \cheasepyinline{iterdbfpath}, takes the path to the \textit{ITERDB} file which \textbf{must be} provided by the user. If the second argument, \cheasepyinline{setParam}, is an empty dictionary, the \cheasepyinline{read\_iterdb} function returns all quantities projected on the original grid ($\rho_{\phi_N}$) provided in the \textit{EXPTNZ} file, otherwise, a source for $\rho_{\psi_N}$ and $\rho_{\phi_N}$ should be provided in the \cheasepyinline{**kwargs} input argument, hence setting \textbf{nrhopsi} in \cheasepyinline{setParam} to 0 will return all the profiles interpolated into the new $\rho_{\psi_N}$ grid provided in the source file, however all the profiles will be interpolated into the new $\rho_{\phi_N}$ counterpart if \textbf{nrhopsi} is set to 1, and \cheasepyinline{read\_iterdb} function will return only the new $\rho_{\psi_N}$ and $\rho_{\phi_N}$ grids to the calling function/script and ignore the original grids provided in the \textit{ITERDB} file. The source of new grids are passed to the \cheasepyinline{**kwargs} input argument of the \cheasepyinline{read\_iterdb} function as described in section(\ref{readProfiles}).\\~\\
In addition to the input profiles, \cheasepyinline{read\_iterdb} function calculates and returns the total pressure. The total pressure calculated in the \cheasepyinline{read\_iterdb} doesn't include the contribution of the fast ions (if any).\\~\\
The \cheasepyinline{read\_iterdb} function returns all the profiles encapsulated in a dictionary in SI units. The name (=key) and description of the returned profiles are listed in the following table:
\begin{center}
	\small
	\begin{tabular}{|c | l |}
		\hline
		\rowcolor{headcolor}
		Measurement & Description \\
		\hline
		\rowcolor{rowcolor}
		Te & Electron Temperature \\
		\rowcolor{rowcolor}
		Ti & Ion Temperature \\
		\rowcolor{rowcolor}
		ne & Electron Density \\
		\rowcolor{rowcolor}
		ni & Ion Density \\
		\rowcolor{rowcolor}
		nz & Impurities Density \\
		\rowcolor{rowcolor}
		Zeff & Effective Atomic Number \\
		\rowcolor{rowcolor}
		Vtor & Toroidal Velocity \\
		\rowcolor{rowcolor}
		rhopsi & $\rho_{\psi_N}$ (with grid source) \\
		\rowcolor{rowcolor}
		rhotor & $\rho_{\phi_N}$ \\
		\rowcolor{rowcolor}
		pprime & Total Pressure Gradient \\
		\rowcolor{rowcolor}
		pressure & Total Pressure of All Species \\
		\hline
	\end{tabular}
\end{center}

\subsubsection{Reading `EXPEQ' File}\label{readEXPEQ}
The \textit{EXPEQ} file is an INPUT/OUTPUT file for \textit{CHEASE} that provides the equilibrium pressure and current profiles on a specific grid. The \textit{EXPEQ} file also provides the equilibrium geometry, such as the (r,z) components of the last closed magnetic surface.\footnote{This magnetic surface should not include any singular point, i.e. x-point, not to cause a problem in finding a state of MHD equilibrium in \textit{CHEASE} code, or an incomplete simulation with an exit error.} All profiles and geometries provided in \textit{EXPEQ} file are `\textit{normalized}' using the equations in section(\ref*{Normalize}).\\~\\
Because of the variety of profiles and grid types that can be used in the \textit{EXPEQ} file, each profile and grid type is assigned a numeric value to refer to it. These numbers are written in the two lines after the list of (r,z) coordinates in the \textit{EXPEQ} file. The first number in the first line specifies the number of grid points of the $\rho_{\psi_N}$ or $\rho_{\phi_N}$ coordinates. The second number in the first line is `\textit{NPPFUN}' and it takes \textcolor{deepgreen}{8} for \textit{pressure} or \textcolor{deepgreen}{4} for \textit{pressure gradient} depending on which one is included in the \textit{EXPEQ} file. The first number in the second line is `\textit{NSTTP}' and it takes \textcolor{deepgreen}{1} for the \textit{current flux density ($ff'$)}, \textcolor{deepgreen}{2} for \textit{surface current ($I^*$)}, \textcolor{deepgreen}{3} for \textit{parallel current ($I_{||}$)}, or \textcolor{deepgreen}{4} for \textit{parallel current density ($J_{||}$)}. The second number in the second line is `\textit{NRHOMESH}' and it takes \textcolor{deepgreen}{0} for \textit{rhopsi ($\rho_{\psi_N}$)} or \textcolor{deepgreen}{1} for \textit{rhotor ($\rho_{\phi_N}$)}.\\~\\
\textit{CheasePy} provides the \cheasepyinline{read\_expeq(expeqfpath,setParam,**kwargs)} function to read the \textit{EXPEQ} file.This function takes the following arguments:
\begin{center}
	\small
	\begin{tabular}{|c | c | l |}
		\hline
		\rowcolor{headcolor}
		Argument & Default & Description \\
		\hline
		\rowcolor{rowcolor}
		expeqfpath & user-input & The path to the \textit{EXPEQ} file. \\
		\rowcolor{rowcolor}
		setParam & \{\} & Select between nrhopsi=0 ($\rho_{\psi_N}$) and nrhopsi=1 ($\rho_{\phi_N}$) grid. \\
		\rowcolor{rowcolor}
		**kwargs & None & Choose a source for the $\rho_{\psi_N}$ and $\rho_{\phi_N}$ grid to interpolate on.\\
		\hline
	\end{tabular}
\end{center}
The first argument, \cheasepyinline{expeqfpath}, takes the path to the \textit{EXPEQ} file which \textbf{must be} provided by the user. If the second argument, \cheasepyinline{setParam}, is an empty dictionary, the \cheasepyinline{read\_expeq} function returns all quantities projected on the original grid ($\rho_{\psi_N}$ or $\rho_{\phi_N}$) provided in the \textit{EXPEQ} file, otherwise, a source for $\rho_{\psi_N}$ and $\rho_{\phi_N}$ should be provided in the \cheasepyinline{**kwargs} input argument, hence setting \textbf{nrhopsi} in \cheasepyinline{setParam} to 0 will return all the profiles interpolated into the new $\rho_{\psi_N}$ grid provided in the source file, however all the profiles will be interpolated into the new $\rho_{\phi_N}$ counterpart if \textbf{nrhopsi} is set to 1, and \cheasepyinline{read\_expeq} function will return only the new $\rho_{\psi_N}$ and $\rho_{\phi_N}$ grids to the calling function/script and ignore the original grids provided in the \textit{EXPEQ} file. The source of new grids are passed to the \cheasepyinline{**kwargs} input argument of the \cheasepyinline{read\_expeq} function as described in section(\ref{readProfiles}).\\~\\
The \cheasepyinline{read\_expeq} function returns all the `\textbf{\textit{normalized}}' pressure profiles and equilibrium geometry encapsulated in a dictionary. The name (=key) and description of the returned profiles are listed in the following table:
\begin{center}
	\small
	\begin{tabular}{|c | l |}
		\hline
		\rowcolor{headcolor}
		Measurement & Description \\
		\hline
		\rowcolor{rowcolor}
		epsilon & Inverse aspect ratio \\
		\rowcolor{rowcolor}
		zgeom & Geometric mean of z-mesh \\
		\rowcolor{rowcolor}
		pedge & Pressure at the Edge \\
		\rowcolor{rowcolor}
		nRZmesh & Size of (r,z) mesh \\
		\rowcolor{rowcolor}
		nrhomesh & Size of $\rho_{\psi_N}$ or $\rho_{\phi_N}$ grid \\
		\rowcolor{rowcolor}
		nrhotype & 0 = $\rho_{\psi_N}$ or 1 = $\rho_{\phi_N}$ \\
		\rowcolor{rowcolor}
		nsttp & 1 = $ff'$, 2 = $I^*$, 3 = $I_{||}$, and 4 = $J_{||}$ \\
		\rowcolor{rowcolor}
		nppfun & 8 = $P$ or 4 = $P'$ \\
		\rowcolor{rowcolor}
		rhopsi & $\rho_{\psi_N}$ (if nrhotype = 0) \\
		\rowcolor{rowcolor}
		rhotor & $\rho_{\phi_N}$ (if nrhotype = 1) \\
		\rowcolor{rowcolor}
		pprime & Total Pressure Gradient (if nppfun = 4) \\
		\rowcolor{rowcolor}
		pressure & Total Pressure of All Species (if nppfun = 8) \\
		\rowcolor{rowcolor}
		ffprime & current flux density (if nsttp = 1) \\
		\rowcolor{rowcolor}
		Istr & surface current (if nsttp = 2) \\
		\rowcolor{rowcolor}
		Iprl & parallel current (if nsttp = 3) \\
		\rowcolor{rowcolor}
		Jprl & parallel current density (if nsttp = 4) \\
		\hline
	\end{tabular}
\end{center}
It worth mention that if the value of `\textit{NPROPT}' parameter in the \textit{chease\_namelist} file is set to a different value of the `\textit{NSTTP}', \textit{CHEASE} uses the current profile that corresponds to the `\textit{NPROPT}' number in the output \textit{EXPEQ} file instead of the one corresponds to the `\textit{NSTTP}'. In addition, the sign of the `\textit{NPROPT}' number is flipped to negative if `\textit{NPPFUN}' is set to \textcolor{deepgreen}{8} and kept positive otherwise. A different value for `\textit{NPROPT}' from its `\textit{NSTTP}' counterpart is needed to run \textit{CHEASE} in an iterative mode with a different input current profile. Hence, we make \textit{CHEASE} to output the desired current profile in the output \textit{EXPEQ} file by setting `\textit{NPROPT}' to the number corresponding that current profile.

\subsubsection{Reading `EQDSK' File}\label{readEQDSK}
The \textit{EQDSK} file is an INPUT/OUTPUT file for \textit{CHEASE} that provides the equilibrium pressure, such as the pressure ($P(\psi)$) and pressure gradient (${dP}/{d\psi}$), and equilibrium current, such as the current flux ($ff'=f{df}/{d\psi}$). It also provides the equilibrium geometry, such as the (r,z) components of the last closed magnetic surface, the safety factor (q) profile, the poloidal mesh ($\psi(R,Z)$), and enough information to construct the $\psi_N$, $\phi_N$, $\rho_{\psi_N}$ and $\rho_{\phi_N}$ grids. All equilibrium profiles and geometries provided in the \textit{EQDSK} file are given in the SI units.\\~\\
\textit{CheasePy} provides the \cheasepyinline{read\_eqdsk(eqdskfpath,setParam,Normalized,**kwargs)} function to read the \textit{EQDSK} file.This function takes the following arguments:
\begin{center}
	\small
	\begin{tabular}{|c | c | l |}
		\hline
		\rowcolor{headcolor}
		Argument & Default & Description \\
		\hline
		\rowcolor{rowcolor}
		eqdskfpath & user-input & The path to the \textit{EQDSK} file. \\
		\rowcolor{rowcolor}
		setParam & \{\} & Select between nrhopsi=0 ($\rho_{\psi_N}$) and nrhopsi=1 ($\rho_{\phi_N}$) grid.\\
		\rowcolor{rowcolor}
		Normalized & False & Return quantities in SI units (False) or Normalized (True). \\
		\rowcolor{rowcolor}
		**kwargs & None & Choose a source for the $\rho_{\psi_N}$ and $\rho_{\phi_N}$ grid to interpolate on.\\
		\hline
	\end{tabular}
\end{center}
The first argument, \cheasepyinline{eqdskfpath}, takes the path to the \textit{EQDSK} file which \textbf{must be} provided by the user. If the second argument, \cheasepyinline{setParam}, is an empty dictionary, the \cheasepyinline{read\_eqdsk} function returns all quantities projected on the original $\phi_N$ grid provided in the \textit{EQDSK} file, otherwise, a source for $\psi_N$ and $\phi_N$ should be provided in the \cheasepyinline{**kwargs} input argument, hence setting \textbf{nrhopsi} in \cheasepyinline{setParam} to 0 will return all the profiles interpolated into the new $\psi_N$ grid provided in the source file, however all the profiles will be interpolated into the new $\phi_N$ counterpart if \textbf{nrhopsi} is set to 1, and \cheasepyinline{read\_eqdsk} function will return only the new $\psi_N$, $\phi_N$, $\rho_{\psi_N}$, and $\rho_{\phi_N}$ grids to the calling function/script and ignore the original grids provided in the \textit{EQDSK} file. The source of new grids are passed to the \cheasepyinline{**kwargs} input argument of the \cheasepyinline{read\_eqdsk} function as described in section(\ref{readProfiles}).\\~\\
The \cheasepyinline{read\_eqdsk} function returns all the profiles and equilibrium geometry encapsulated in a dictionary. The name (=key) and description of the returned profiles are listed in the following table:\footnote{Due to the long list of parameters returned from the \cheasepyinline{read\_eqdsk} function after reading the \textit{EQDSK} file, we ONLY listed here those are used frequently by \textit{CheasePy} code. The \cheasepyinline{read\_efit\_file} function has a full list of all parameters that are read from the \textit{EQDSK} file. This \textcolor{blue}{\href{https://w3.pppl.gov/ntcc/TORAY/G_EQDSK.pdf}{link}} has a full description of the \textit{EQDSK} parameters.}
\begin{center}
	\small
	\begin{tabular}{|c | l |}
		\hline
		\rowcolor{headcolor}
		Measurement & Description \\
		\hline
		\rowcolor{rowcolor}
		q & Safety Factor Profile \\
		\rowcolor{rowcolor}
		BCTR & Toroidal B at Magnetic Axis \\
		\rowcolor{rowcolor}
		RCTR & R-Coordinate at Magnetic Axis \\
		\rowcolor{rowcolor}
		CURNT & Total Experimental Toroidal Current \\
		\rowcolor{rowcolor}
		PSIN & $\psi_N$ \\
		\rowcolor{rowcolor}
		PHIN & $\phi_N$ (with grid source) \\
		\rowcolor{rowcolor}
		rhopsi & $\rho_{\psi_N}$ \\
		\rowcolor{rowcolor}
		rhotor & $\rho_{\phi_N}$ (with grid source) \\
		\rowcolor{rowcolor}
		rbound & r-coordinate of magnetic boundary surface \\
		\rowcolor{rowcolor}
		zbound & z-coordinate of magnetic boundary surface \\
		\rowcolor{rowcolor}
		pprime & Total Pressure Gradient ($P'(\psi)$) \\
		\rowcolor{rowcolor}
		pressure & Total Pressure of All Species ($P(\psi)$) \\
		\rowcolor{rowcolor}
		ffprime & current flux function ($ff'(\psi)$) \\
		\hline
	\end{tabular}
\end{center}

\subsubsection{Reading `CHEASE' File}\label{readCHEASE}
The \textit{CHEASE} file is an OUTPUT file for \textit{CHEASE} that provides the density and temperature for electrons, ions, and impurities, equilibrium pressure and currents, equilibrium geometry, safety factor (q) profile, and polodial and toroidal grids. All equilibrium profiles and geometries provided in the \textit{CHEASE} file are given in the SI units.\\~\\
\textit{CheasePy} provides the \cheasepyinline{read\_chease(cheasefpath,setParam,Normalized,**kwargs)} function to read the \textit{CHEASE} file. This function takes the following arguments:
\begin{center}
	\small
	\begin{tabular}{|c | c | l |}
		\hline
		\rowcolor{headcolor}
		Argument & Default & Description \\
		\hline
		\rowcolor{rowcolor}
		cheasefpath & user-input & The path to the \textit{CHEASE} file. \\
		\rowcolor{rowcolor}
		setParam & \{\} & Select between nrhopsi=0 ($\rho_{\psi_N}$) and nrhopsi=1 ($\rho_{\phi_N}$) grid.\\
		\rowcolor{rowcolor}
		Normalized & False & Return quantities in SI units (False) or Normalized (True). \\
		\rowcolor{rowcolor}
		**kwargs & None & Choose a source for the $\rho_{\psi_N}$ and $\rho_{\phi_N}$ grid to interpolate on.\\
		\hline
	\end{tabular}
\end{center}
The first argument, \cheasepyinline{cheasefpath}, takes the path to the \textit{CHEASE} file which \textbf{must be} provided by the user. If the second argument, \cheasepyinline{setParam}, is an empty dictionary, the \cheasepyinline{read\_chease} function returns all quantities projected on the original $\phi_N$ grid provided in the \textit{CHEASE} file, otherwise, a source for $\psi_N$ and $\phi_N$ should be provided in the \cheasepyinline{**kwargs} input argument, hence setting \textbf{nrhopsi} in \cheasepyinline{setParam} to 0 will return all the profiles interpolated into the new $\psi_N$ grid provided in the source file, however all the profiles will be interpolated into the new $\phi_N$ counterpart if \textbf{nrhopsi} is set to 1, and \cheasepyinline{read\_chease} function will return only the new $\psi_N$, $\phi_N$, $\rho_{\psi_N}$, and $\rho_{\phi_N}$ grids to the calling function/script and ignore the original grids provided in the \textit{CHEASE} file. The source of new grids are passed to the \cheasepyinline{**kwargs} input argument of the \cheasepyinline{read\_chease} function as described in section(\ref{readProfiles}).\\~\\
The \cheasepyinline{read\_chease} function returns all the profiles and equilibrium geometry encapsulated in a dictionary. The name (=key) and description of the returned profiles are listed in the following table:\footnote{Due to the long list of parameters returned from the \cheasepyinline{read\_chease} function after reading the \textit{CHEASE} file, we ONLY listed here those are frequently used by \textit{CheasePy} code. The \cheasepyinline{read\_chease} function has a full list of all parameters that are read from the \textit{CHEASE} file. This \textcolor{blue}{\href{https://www.sciencedirect.com/science/article/pii/001046559600046X}{paper}} has a full description of the \textit{CHEASE} parameters.}
\begin{center}
	\small
	\begin{tabular}{|c | l |}
		\hline
		\rowcolor{headcolor}
		Measurement & Description \\
		\hline
		\rowcolor{rowcolor}
		q & Safety Factor Profile \\
		\rowcolor{rowcolor}
		B0EXP & Toroidal B at Magnetic Axis \\
		\rowcolor{rowcolor}
		R0EXP & R-Coordinate at Magnetic Axis \\
		\rowcolor{rowcolor}
		ITEXP & Total Experimental Toroidal Current \\
		\rowcolor{rowcolor}
		PSIN & $\psi_N$ \\
		\rowcolor{rowcolor}
		PHIN & $\phi_N$ (with grid source) \\
		\rowcolor{rowcolor}
		rhopsi & $\rho_{\psi_N}$ \\
		\rowcolor{rowcolor}
		rhotor & $\rho_{\phi_N}$ (with grid source) \\
		\rowcolor{rowcolor}
		rbound & r-coordinate of magnetic boundary surface \\
		\rowcolor{rowcolor}
		zbound & z-coordinate of magnetic boundary surface \\
		\rowcolor{rowcolor}
		shear & Magnetic Shear ($\tilde{s}(\psi)$) \\
		\rowcolor{rowcolor}
		signeo & Neoclassical Conductivity ($\sigma_{neo}(\psi)$) \\
		\rowcolor{rowcolor}
		Te (Ti) & Electron (Ion) Temperature ($T_e(\psi)$ ($T_i(\psi)$)) \\
		\rowcolor{rowcolor}
		ne (ni,nz) & Electron (Ion,impurities) Density ($n_e(\psi)$ ($n_i(\psi)$,$n_z(\psi)$)) \\
		\rowcolor{rowcolor}
		pprime & Total Pressure Gradient ($P'(\psi)$) \\
		\rowcolor{rowcolor}
		pressure & Total Pressure of All Species ($P(\psi)$) \\
		\rowcolor{rowcolor}
		ffprime & current flux function ($ff'(\psi)$) \\
		\rowcolor{rowcolor}
		Ibs & Bootstrap Current ($I_{bs}(\psi)$) \\
		\rowcolor{rowcolor}
		Jbs & Bootstrap Current Density ($J_{bs}(\psi)$) \\
		\rowcolor{rowcolor}
		Iprl & Parallel Current ($I_{||}(\psi)$) \\
		\rowcolor{rowcolor}
		Jprl & Parallel Current Density ($J_{||}(\psi)$) \\
		\rowcolor{rowcolor}
		Istr & Surface Current ($I^*(\psi)$) \\
		\rowcolor{rowcolor}
		Itor & Toroidal Current ($I_{\phi}(\psi)$) \\
		\rowcolor{rowcolor}
		Jtor & Toroidal Current Density ($J_{\phi}(\psi)$) \\
		\hline
	\end{tabular}
\end{center}
There are also a complete list of the magnetic coefficients ($g_{ij}$), Jacobian transformation matrix ($J$), and other parameters that can be of interested to the users.

\newpage
\subsection{Creating Input Files}
Before any iteration of \textit{CheasePy} the \cheasepyinline{write\_exptnz} and \cheasepyinline{write\_expeq} functions are called to create the \textit{EXPTNZ} and \textit{EXPEQ} files, respectively, which are used as input files to \textit{CHEASE} code. However, if the \textit{namelist} parameter \textit{NEQDSK} is set to \textit{zero}, only the \cheasepyinline{write\_expntz} function is called and the \textit{EQDSK} file is used as input to \textit{CHEASE}.\footnote{\textit{CheasePy} renames \textit{EQDSK} file to \textit{EXPEQ} to be recognized by \textit{CHEASE} code.} The user can specify the sources of each profile to be used to create the \textit{EXPTNZ} and \textit{EXPEQ} files in the \cheasepyinline{setParam} input argument to the \cheasepyinline{write\_exptnz} and \cheasepyinline{write\_expeq} functions, as we explained in sections \ref{createEXPTNZ} and \ref{createEXPEQ}. After every iteration, the output files are renamed by adding a suffix \textit{\_iterxxx} to reflect the iteration number, for instance \textit{chease\_iter000.h5}, \textit{EXPEQ\_iter000.OUT}, and \textit{EXPTNZ\_iter000.OUT} for the \textit{zeroth} iteration.
\subsubsection{Creating EXPEQ File}\label{createEXPEQ}
The \textit{EXPEQ} file has the (r,z) coordinates of the LCMS, in addition to the pressure and current profiles. \textit{CheasePy} provides the \cheasepyinline{write\_expeq(setParam,outfile,**kwargs)} function to create the \textit{EXPEQ} file and/or return all the contents of that file as a dictionary structure with field names that follow the description in section(\ref{readEXPEQ}). This function takes the following arguments:
\begin{center}
	\small
	\begin{tabular}{|c | c | l |}
		\hline
		\rowcolor{headcolor}
		Argument & Default & Description \\
		\hline
		\rowcolor{rowcolor}
		setParam & \{\} & Specify the sources for different profiles.\\
		\rowcolor{rowcolor}
		outfile & True & Create an output EXPEQ file. \\
		\rowcolor{rowcolor}
		**kwargs & None & Specify the path to the source files.\\
		\hline
	\end{tabular}
\end{center}
The first argument, \cheasepyinline{setParam}, has a dictionary structure to specify the types and sources of the grid (\cheasepyinline{nrhomesh}), the pressure profile (\cheasepyinline{nppfun}), and current profile (\cheasepyinline{nsttp}), each of these inputs takes a list that has the structure of \textcolor{deepgreen}{[type,source]}. There are three sources accepted in \cheasepyinline{setParam} for \cheasepyinline{nrhomesh}: \{0,`chease'\}, \{1,`eqdks'\}, or \{7,`imported'\}, six sources for \cheasepyinline{nppfun}: \{0,`chease'\}, \{1,`eqdks'\}, \{2,`expeq'\}, \{3,`exptnz'\}, \{4,`profiles'\}, \{5,`iterdb'\}, or \{7,`imported'\}, and four sources for \cheasepyinline{nsttp}: \{0,`chease'\}, \{1,`eqdks'\}, \{2,`expeq'\}, or \{7,`imported'\}. The path to the source files, i.e. `chease', `eqdsk', `expeq', `exptnz', `profiles', and `iterdb', must be provided in the \cheasepyinline{**kwargs} argument. However, if the \{7,`imported'\} sources is chosen, a dictionary of the profiles (\{name,data\}) must be provided in the \cheasepyinline{**kwargs} argument, e.g. \{`pressure',[list of pressure data]\}. All the returned profiles are projected (interpolated) into the grid provided in the \cheasepyinline{nrhomesh} source by setting the grid source in the \cheasepyinline{setParam} in the read function to the common grid source.\\~\\
On the other hand, the type of the grid (\cheasepyinline{nrhomesh}) is either \{0,`rhopsi'\} or \{1,`rhoto'\}, the type of the pressure profile (\cheasepyinline{nppfun}) is either \{4,`pprime'\} or \{8,`pressure'\}, and the type of the current profile (\cheasepyinline{nsttp}) is either \{1,`ffprime'\}, \{2,`istr'\}, \{3,`iprl'\}, or \{4,`jprl'\}. The \textit{CheasePy} user has to ensure that the source file has the requested profile. For example, the \textit{EQDSK} file has a single current profile, i.e. \textit{ffprime} and both pressure profiles, i.e. \textit{pressure} and \textit{pprime}. The current profile in the \textit{EXPEQ} file depends on the value of \textit{NPROPT} parameter in the input \textit{chease\_namelist} file. However, the \textit{CHEASE} source has all types of profiles. Finally, the user should remember that all quantities in the \textit{EXPEQ} source are normalized, which is not the case in other sources.\\~\\
If the \cheasepyinline{setParam} is empty dictionary, the types are set to their default values, i.e. \cheasepyinline{nrhomesh} = 0, \cheasepyinline{nppfun} = 8, and \cheasepyinline{nsttp} = 1. Hence, any provided source in the \cheasepyinline{**kwargs} will be used to create the new \textit{EXPEQ} file, otherwise, \textit{CheasePy} will search for the \textit{EQDSK} and use it as a source for the profiles. In a similar way, the source for the (r,z) coordinates of the LCMS is taken to be the same source for the \cheasepyinline{nrhomesh} grid.\\~\\
The second input argument (outfile) takes \textit{True} if the user wants to create the \textit{EXPEQ} file and receive a return dictionary with the content of the \textit{EXPEQ} file, or \textit{False} if the user wants only that returned dictionary without creating the \textit{EXPEQ} file.

\subsubsection{Creating EXPTNZ File}\label{createEXPTNZ}
The \textit{EXPTNZ} file contains the density and temperature profiles for electrons and ions, and the effective atomic number. \textit{CheasePy} provides the \cheasepyinline{write\_exptnz(setParam,outfile,**kwargs)} function to create the \textit{EXPTNZ} file and/or return all the contents of that file as a dictionary structure with field names that follow the description in section(\ref{readEXPTNZ}). This function takes the following arguments:
\begin{center}
	\small
	\begin{tabular}{|c | c | l |}
		\hline
		\rowcolor{headcolor}
		Argument & Default & Description \\
		\hline
		\rowcolor{rowcolor}
		setParam & \{\} & Specify the sources for different profiles.\\
		\rowcolor{rowcolor}
		outfile & True & Create an output EXPEQ file. \\
		\rowcolor{rowcolor}
		**kwargs & None & Specify the path to the source files.\\
		\hline
	\end{tabular}
\end{center}
The first argument, \cheasepyinline{setParam}, has a dictionary structure to specify the types and sources of the grid (\cheasepyinline{nrhomesh}), the electron density and temperature profiles (\cheasepyinline{eprofiles}), and ion density and temperature profiles (\cheasepyinline{iprofiles}). There are three sources accepted in \cheasepyinline{setParam} for \cheasepyinline{nrhomesh}: \{0,`chease'\}, \{1,`eqdks'\}, or \{7,`imported'\}, four sources for \cheasepyinline{eprofiles} and \cheasepyinline{iprofiles}: \{0,`chease'\}, \{3,`exptnz'\}, \{4,`profiles'\}, \{5,`iterdb'\}, and/or \{7,`imported'\}.  The path to the source files, i.e. `chease', `eqdsk', `exptnz', `profiles', and `iterdb', must be provided in the \cheasepyinline{**kwargs} argument. However, if the \{7,`imported'\} sources is chosen, a dictionary of the profiles (\{name,data\}) must be provided in the \cheasepyinline{**kwargs} argument, e.g. \{`pressure',[list of pressure data]\}. All the returned profiles are projected (interpolated) into the grid provided in the \cheasepyinline{nrhomesh} source by setting the grid source in the \cheasepyinline{setParam} in the read function to the common grid source. On the other hand, the type of the grid (\cheasepyinline{nrhomesh}) is either \{0,`rhopsi'\} or \{1,`rhoto'\}.\\~\\
If the \cheasepyinline{setParam} is empty dictionary, the type of the grid is set to its default value, i.e. \cheasepyinline{nrhomesh} = 0. Hence, any provided source will be used to create the \textit{EXPTNZ} file, otherwise, the \textit{EXPTNZ} will be the source of the profiles and grid. Finally, the user should remember that all quantities in the \textit{EXPTNZ} source are in SI units.\\~\\
The second input argument (outfile) takes \textit{True} if the user wants to create the \textit{EXPTNZ} file and receive a return dictionary with the content of the \textit{EXPTNZ} file, or \textit{False} if the user wants only that returned dictionary without creating the \textit{EXPTNZ} file.

\newpage
\subsection{Running CheasePy}
A user can run \textit{CheasePy} code in a directory that has \textit{CHEASE} code executable named \textit{chease\_hdf5} and \textit{runchease.py} by typing this command in a terminal window:
\begin{center}
\cheasepyinline{python runchease.py}
\end{center}
This command calls the \cheasepyinline{cheasepy} function to prepare the required input files and run \textit{CHEASE} code. The \cheasepyinline{cheasepy} function takes \textit{four} input arguments, i.e. \cheasepyinline{srcVals}, \cheasepyinline{namelistVals}, \cheasepyinline{cheaseVals}, and \cheasepyinline{importedVals}, that set the sources of density, temperature, and pressure profiles, geometry parameters, the list of \textit{CHEASE} namelist parameters, and the \textit{CheasePy} mode of operation. Details for these input arguments are described below.
\subsubsection{Profiles and Geometry Sources}
In the \textit{runchease.py} script a dictionary variable named \cheasepyinline{srcVals} is initialized, and then the sources are specified for each type. The temperature and density profiles for the ions, electrons, and impurities can be taken from different sources, such as \textit{profiles}, \textit{exptnz}, \textit{iterdb}, or \textit{chease} files, with the source for electrons and ions+impurities profiles can be different. The density and temperature of electrons, ions, and impurities (if any) taken from any combinations of the previous sources can also be used to calculate the total pressure. Otherwise, the total pressure, the current, and equilibrium geometry are taken directly from \textit{eqdsk}, \textit{expeq}, or \textit{chease} files.\\~\\
With These different sources, each source is identified by its name or a number as shown in the table below:
\begin{center}
	\begin{tabular}{| c | c | c | c | c | c | c | c |}
		\hline
		\cellcolor{headcolor} Number & \cellcolor{rowcolor}0 & \cellcolor{rowcolor}1 & \cellcolor{rowcolor}2 & \cellcolor{rowcolor}3 & \cellcolor{rowcolor}4 & \cellcolor{rowcolor}5 & \cellcolor{rowcolor}7 \\
		\hline
		\cellcolor{headcolor} Source & \cellcolor{rowcolor}chase & \cellcolor{rowcolor}eqdsk & \cellcolor{rowcolor}expeq & \cellcolor{rowcolor}exptnz & \cellcolor{rowcolor}profiles & \cellcolor{rowcolor}iterdb & \cellcolor{rowcolor}imported \\
		\hline
	\end{tabular}
\end{center}
\begin{example}\label{exSetSources}
If the sources for electron profile, ion profile, pressure, and current are \textit{exptnz}, \textit{profiles},  \textit{eqdsk}, and \textit{expeq} files, respectively. Also, all the profiles will be interpolated on the $\rho_{\phi_N}$ defined in \textit{eqdsk} file. Hence, \cheasepyinline{srcVals} variable will be defined as follow:
\begin{cheasepy}
	srcVals = {}
	srcVals['iprofiles_src'] = 'profiles'
	srcVals['eprofiles_src'] = 'exptnz'
	srcVals['pressure_src'] = 'eqdsk'
	srcVals['current_src'] = 'expeq'
	srcVals['rhomesh_src'] = 'eqdsk'
\end{cheasepy}
\end{example}
\begin{example}\label{exSetRhoSrc2None}
If the sources for electron and ion profiles are taken from the \textit{profiles} file, however, the pressure and current profiles are taken from the \textit{eqdsk} file. The $\rho_{\phi_N}$ grid is taken from the common source of each related profile, i.e. the $\rho_{\phi_N}$ grid used in the generated \textit{exptnz} file is the same as that in the \textit{profiles} file, and the $\rho_{\phi_N}$ grid used in the generated \textit{expeq} is the same as that in the \textit{eqdsk} file. Hence, \cheasepyinline{srcVals} variable will be defined as follow:
\begin{cheasepy}
	srcVals = {}
	srcVals['iprofiles_src'] = 'profiles'
	srcVals['eprofiles_src'] = 'profiles'
	srcVals['pressure_src'] = 'eqdsk'
	srcVals['current_src'] = 'eqdsk'
	srcVals['rhomesh_src'] = None
\end{cheasepy}
\end{example}
In addition to the profiles and grid sources defined in the \cheasepyinline{srcVals} variable, there is another variable also defined in the \cheasepyinline{srcVals} which is not a source of any profile but a method of how to treat the last closed magnetic surface (LCMS). This variable is called \cheasepyinline{boundary\_type}, and it takes {\ttb\color{deepgreen}{0}} or \cheasepyinline{'asis'} in case the LCMS will be written as it is read from the source file without any massaging or processing of its coordinates. However, if \cheasepyinline{boundary\_type} receives {\ttb\color{deepgreen}{1}} or \cheasepyinline{'interp'} the coordinates taken from the source file will be interpolated to have the LCMS defined at a specific length of $\rho_{\psi_N}$. This interpolation is needed when the LCMS has an x-point (i.e. singular point) which makes \textit{CHEASE} either crashes or gives irrelevant/incorrect results in case the \textit{TENSBND} parameter wasn't set to the right value, i.e. -1.\footnote{Although \textit{TENSBND} is a very good tool provided by \textit{CHEASE} to interpolate around the x-point, the code user doesn't have any control on how close the new LCMS will be close to the original one. However, using the \cheasepyinline{'interp'} attribute with \cheasepyinline{boundary\_type} allows the user to set a cutoff value for $\rho_{\psi_N}$ that defines the LCMS.}\\~\\
The interpolation function \cheasepyinline{magsurf\_solvflines} is defined in the \textit{efittools.py} python package, and it returns the (r,z) coordinates of the LCMS after solving the magnetic field lines equations using a fifth-order Rung-Kutta method. The \cheasepyinline{magsurf\_solvflines} function receives the path to the \textit{eqdsk} file in the \cheasepyinline{eqdskfpath} argument, a cutoff $\psi_N$ value for the LCMS in the \cheasepyinline{psi}, and a tolerance value for the desired accuracy in the \cheasepyinline{eps} argument. When the \textit{NEQDSK} in \cheasepyinline{chease\_namelist} parameter is set to {\ttb\color{deepgreen}{0}}, the function \cheasepyinline{magsurf\_solvflines} is called with \cheasepyinline{psi} {\ttb\color{deepgreen}{0.999}}.\footnote{The value of \cheasepyinline{psi} in the call to the \cheasepyinline{magsurf\_solvflines} can be increased or decreased by \textit{CheasePy} user. However, with a \cheasepyinline{psi} value that is very cloase to {\ttb\color{deepgreen}{1.0}} the magnetic field lines solver might return (r,z) coordinates of an open field surface (outside the separatix), which causes \textit{CHEASE} code to crash. So the \textit{CheasePy} user has to verify that a closed surface is returned from \cheasepyinline{magsurf\_solvflines} function by using a suitable \cheasepyinline{psi} value.}\\~\\
In the table of profile and grid sources above we referred to a source for the profiles called {\ttb\color{deepgreen}{imported}}. This source type is usefule when \textit{CheasePy} user wants to import a profile for any quantity from an external source, or the user wants to do some processing on any profile's data before include it as an input to \textit{CHEASE} code. All imported profiles which are passed to \cheasepyinline{cheasepy} function through the \cheasepyinline{importedVals} argument must be accompanied with the $\rho_{\psi_N}$ and $\rho_{\phi_N}$ grids. Passing the imported profiles along with their corresponding grid is required because these grids are used for interpolating those profiles onto the common grid on each input file to \textit{CHEASE} code, i.e. \textit{exptnz} and \textit{expeq}. This means if the user has profiles from several external sources, all of these profiles have to be projected onto the same grids before passing them to the \cheasepyinline{cheasepy} function.
\begin{example}\label{exImportAll}
A file, externalSource.dat, has four columns for $\rho_{\psi_N}$, $\rho_{\phi_N}$, $n_e$, and $T_e$ profiles, and a user need to use these profiles as input to \textit{CHEASE}. The steps the user has to follow to use these profiles from that external source to \textit{CHEASE} are as follow:
\begin{cheasepy}
	import numpy as npy
	eprofiles = npy.loadtxt(externalSource.dat)
	
	importedVals = {}
	importedVals['rhotor'] = npy.array(eprofiles[:,0])
	importedVals['rhopsi'] = npy.array(eprofiles[:,1])
	importedVals['Te']   = npy.array(eprofiles[:,2])*1.0e3
	importedVals['ne']   = npy.array(eprofiles[:,3])*1.0e19
	
	srcVals = {}
	srcVals['iprofiles_src'] = 'exptnz'
	srcVals['eprofiles_src'] = 'imported'
	srcVals['pressure_src'] = 'eqdsk'
	srcVals['current_src'] = 'expeq'
	srcVals['rhomesh_src'] = 'eqdsk'
\end{cheasepy}
\end{example}
In example(\ref{exImportAll}) the electron density and temperature profiles, and the $\rho_{\psi_N}$ and $\rho_{\phi_N}$ grids are assigned to different keys in the \cheasepyinline{importedVals} dictionary. Then, the \cheasepyinline{'eprofiles\_src'} item in the \cheasepyinline{srcVals} dictionary is set to \cheasepyinline{'imported'}. Hence, all the electron profiles will be taken from the \cheasepyinline{'imported'} profiles. However, if the user wants to take only the electron temperature from the external source and take the electron density from the \textit{profiles} file, the user can import only the electron temperature from the external source, and keep the \cheasepyinline{'eprofiles_src'} item in the \cheasepyinline{srcVals} assigned to \cheasepyinline{'profiles'}, as shown in example(\ref{exImportTe}) below.
\begin{example}\label{exImportTe}
A user wants to import only the temperature profile from an external source and take the electron density and ion profiles from the \textit{profiles} file.
\begin{cheasepy}
	import numpy as npy
	eprofiles = npy.loadtxt(externalSource.dat)
	
	importedVals = {}
	importedVals['rhotor'] = npy.array(eprofiles[:,0])
	importedVals['rhopsi'] = npy.array(eprofiles[:,1])
	importedVals['Te']   = npy.array(eprofiles[:,2])*1.0e3
	
	srcVals = {}
	srcVals['iprofiles_src'] = 'profiles'
	srcVals['eprofiles_src'] = 'profiles'
	srcVals['pressure_src'] = 'eqdsk'
	srcVals['current_src'] = 'expeq'
	srcVals['rhomesh_src'] = 'eqdsk'
\end{cheasepy}
\end{example}
The question that might be arisen now is: How will \textit{CheasePy} use the \cheasepyinline{'imported'} electron temperature if the user didn't pass it to the \cheasepyinline{srcVals} dictionary? The answer to this question is simple: When \cheasepyinline{importedVals} dictionary is passed to \cheasepyinline{cheasepy} function as an input argument, it, in turn, passes it to \cheasepyinline{write\_exptnz} and \cheasepyinline{write\_expeq} functions. When any of these functions receives the content of \cheasepyinline{importedVals} dictionary, they inspect that content and enforce any imported profile to override the corresponding profile provided in the \cheasepyinline{srcVals} dictionary. Therefore, the \cheasepyinline{'imported'} electron temperature profile provided in \cheasepyinline{importedVals} overrides the same profile provided in \cheasepyinline{srcVals}. This takes place for any profile used to construct \textit{EXPTNZ} and \textit{EXPEQ} files.\\~\\
As we explained earlier, the pressure profile is compromised of (at least) two components from ions and electrons, where $P_T = n_eT_e + n_iT_i$. In addition, there might be contributions from the impurities and fast ions with the total pressure given by: $P_T = n_eT_e + n_iT_i + n_zT_z + n_bT_b$, where $n_b$ and $T_b$ are the density and temperature of fast ions. A common mistake is done in examples \ref{exImportAll} and \ref{exImportTe} when the pressure profile is taken directly from the \textit{EQDSK} file without considering the changes in one of its components which is the electron density and/or temperature profiles. Therefore, the total pressure has to be recalculated using the contribution of the imported profiles of the electron density and/or temperature. The following example explains how to calculate the new total pressure using the new electron profile.
\begin{example}\label{exImportPt}
A user has two \textit{iterdb} files, \textit{basecase.iterdb} which has the original profiles of the electrons, ions, and impurities, and \textit{modTeprof.iterdb} which has a modified electron temperature ($T_e$) profile. The following code shows how to include the contribution of the modified electron temperature into the pressure profile.
\begin{cheasepy}
	import cheasepy
	
	inParam = {nrhomesh:0}
	
	baseITERDB = cheasepy.read_iterdb(iterdbfpath='basecase.iterdb',
	                                  setParam=inParam,eqdsk='EQDSK')
	modITERDB  = cheasepy.read_iterdb(iterdbfpath='modTeProf.iterdb',
	                                  setParam=inParam,eqdsk='EQDSK')
	
	importedVals = {}
	importedVals['rhotor'] = modITERDB['rhotor']
	importedVals['rhopsi'] = modITERDB['rhopsi']
	importedVals['Te']     = modITERDB['Te']
	
	eqdskdata = cheasepy.read_eqdsk(eqdskfpath='EQDSK')
	
	importedVals['pressure'] = eqdskdata['pressure']-baseITERDB['pressure']
	                                                + modITERDB['pressure']
	
	srcVals = {}
	srcVals['iprofiles_src'] = 'iterdb'
	srcVals['eprofiles_src'] = 'iterdb'
	srcVals['pressure_src'] = 'eqdsk'
	srcVals['current_src'] = 'expeq'
	srcVals['rhomesh_src'] = 'eqdsk'
\end{cheasepy}
\end{example}
In example(\ref{exImportPt}), although the \cheasepyinline{'pressure\_src'} is set to \cheasepyinline{'eqdsk'}, the imported value of the total pressure will override the sourced profile in the \cheasepyinline{write\_expeq} functions. Similarly, the imported electron temperature profile will override the sourced profile in the \cheasepyinline{write\_exptnz} functions.\\~\\
All the profiles that can be externally imported into \textit{CheasePy} an override sourced profiles are listed in the following table:\footnote{This list can be extended to include other imported profiles by considering them in the \cheasepyinline{write\_expeq} and \cheasepyinline{write\_exptnz} functions.}
\begin{center}
\begin{tabular}{| c | l | c | l |}
	\hline
	\rowcolor{headcolor}
	Profile & Description & Profile & Description \\
	\hline
	\rowcolor{rowcolor}
	rhopsi & Poloidal Grid & rhotor & Toroidal Grid \\
	\rowcolor{rowcolor}
	ne & Electron Density & Te & Electron Temperature \\
	\rowcolor{rowcolor}
	ni & Ion Density & Ti & Ion Temperature \\
	\rowcolor{rowcolor}
	nz & Impurity Density & Tz & Impurity Temperature \\
	\rowcolor{rowcolor}
	Iprl & Parallel Current & Jprl & Parallel Current Density \\
	\rowcolor{rowcolor}
	ffprime & Current Flux Density & Istr & Surface Current \\
	\rowcolor{rowcolor}
	pressure & Total Pressure & pprime & Pressure Gradient \\
	\rowcolor{rowcolor}
	Zeff & Effective Atomic Number & ~ & ~ \\
	\hline
\end{tabular}
\end{center}
All the examples and discussions above consider only the sources for input profiles for the zeroth iteration in \textit{CheasePy} which reconstruct the MHD equilibrium depending on the input profiles and equilibrium geometry. If \textit{CheasePy} code runs for several iterations it will keep using the same profiles given in the source files, but ignoring the imported profiles after the zeroth iteration. Hence, another source of the profiles and equilibrium geometry sometimes required to track the changes made in the zeroth iteration by the imported profiles. \textit{CheasePy} allows the user to change the source of each profile from one iteration to another by providing these sources in a Python \textit{list}. For every profile the user needs to change its source, this source has to be of type list and of length equals the number of source that the user needs to switch between them. \textit{CheasePy} code will keep using the last source in the list for the rest of iterations in case the length of the source list is less than the number of iterations. Usually the generated \textit{EXPTNZ}, \textit{EXPEQ}, \textit{EQDSK}, and \textit{CHEASE} files after each iteration have updated profiles and equilibrium geometry.\\~\\
If the user needs to do several iteration for the study case in example(\ref{exImportPt}), the second source of \cheasepyinline{'eprofile\_src'} need to be taken from \textit{EXPTNZ} or \textit{CHEASE} files, however, the \cheasepyinline{'pressure\_src'} need to be taken from \textit{EXPEQ}, \textit{EQDSK}, or \textit{CHEASE} files to consider the imported profiles that was enforced in the zeroth iteration. Hence, example(\ref{exImportPt}) can be rewritten as follow:
\begin{example}\label{exModImportPt}
A user has two \textit{iterdb} files, \textit{basecase.iterdb} which has the original profiles of the electrons, ions, and impurities, and \textit{modTeprof.iterdb} which has a modified electron temperature ($T_e$) profile. The following code shows how to include the contribution of the modified electron temperature into the pressure profile, and how to consider these variations in further iterations. Also, the user needs to change the source of the grid after the zeroth iteration to be taken from the \textit{CHEASE} source file.
\begin{cheasepy}
	import cheasepy
	
	inParam = {nrhomesh:0}
	
	baseITERDB = cheasepy.read_iterdb(iterdbfpath='basecase.iterdb',
	setParam=inParam,eqdsk='EQDSK')
	modITERDB  = cheasepy.read_iterdb(iterdbfpath='modTeProf.iterdb',
	setParam=inParam,eqdsk='EQDSK')
	
	importedVals = {}
	importedVals['rhotor'] = modITERDB['rhotor']
	importedVals['rhopsi'] = modITERDB['rhopsi']
	importedVals['Te']     = modITERDB['Te']
	
	eqdskdata = cheasepy.read_eqdsk(eqdskfpath='EQDSK')
	
	importedVals['pressure'] = eqdskdata['pressure']-baseITERDB['pressure']
													+ modITERDB['pressure']
	
	srcVals = {}
	srcVals['iprofiles_src'] = 'iterdb'
	srcVals['eprofiles_src'] = ['iterdb','exptnz']
	srcVals['pressure_src'] = ['eqdsk','expeq']
	srcVals['current_src'] = 'expeq'
	srcVals['rhomesh_src'] = ['eqdsk','chease']
\end{cheasepy}
\end{example}

\subsubsection{CHEASE Namelist Parameters}
A list of \textit{namelist parameters} is provided in section(\ref{CreateNamelist}). These \textit{namelist parameters} are required for \textit{CHEASE} code to run properly and give relevant results to the case under study. \textit{CheasePy} code has a default value for each parameter in the \textit{namelist}, but the user can change any of them depending on several factors, such as the input profiles, input sources, and desired resolution. \textit{CheasePy} provides a dictionary-type variable called \cheasepyinline{namelistVals} that you can use to assign a new value for any of the parameter in the \textit{namelist}. Then, this \cheasepyinline{namelistVals} is passed as an input argument to the \cheasepyinline{cheasepy} function to create the \textit{chease\_namelist} file using the user-defined and default values for each parameter.\footnote{The \textit{CheasePy} user can revise the list of parameters required by \textit{CHEASE} code in the table in section(\ref{CreateNamelist}).}\\~\\
If the \textit{CheasePy} user runs multiple iterations of the \textit{CHEASE} code and needs to change any of the input parameters from one iteration to another, an equal length list of these parameter values need to be assigned to each item in the \cheasepyinline{namelistVals} input argument, even if the values of these parameters don't change from one iteration to another. In other words, if a user needs to change the type of couple of \textit{namelist parameters} from \cheasepyinline{int} or \cheasepyinline{float} to \cheasepyinline{list} without changing all the type of other \textit{namelist parameters} an error will be raised by \textit{CheasePy} as it expects all parameters in the \textit{namelist} of the same type, i.e \textit{single-valued} or \textit{list}.
\begin{example}
A user needs to run \textit{CHEASE} code using pressure and current profiles from the \textit{EQDSK} for the zeroth iteration and from \textit{EXPEQ} for other iterations. Also, \textit{CHEASE} code should expect $ff'$ in the zeroth iteration, and $I_{||}$ for other iterations.
\begin{cheasepy}
	namelistVals = {}
	namelistVals['NS']        = [128,128]
	namelistVals['NT']        = [128,128]
	namelistVals['NISO']      = [128,128]
	namelistVals['NPSI']      = [512,512]
	namelistVals['NCHI']      = [512,512]
	namelistVals['RELAX']     = [0.0,0.0]
	namelistVals['NEQDSK']    = [1,0]
	namelistVals['TENSBND']   = [0,0]
	namelistVals['TENSPROF']  = [0,0]
	namelistVals['NRHOMESH']  = [0,0]
	namelistVals['cocos_in']  = [1,1]
	namelistVals['cocos_out'] = [11,11]
	
	namelistVals['NSTTP']     = [1,3]
	namelistVals['NPROPT']    = [3,3]
	namelistVals['NPPFUN']    = [8,8]
\end{cheasepy}
\end{example}
It worth reminding the user that there are many other parameters those are set to their default values, and out-of-all the most two important parameters are \textit{NRBOX} and \textit{NZBOX} which specify the resolution of the profiles and equilibrium geometry in the output \textit{EQDSK}.\footnote{The output \textit{EQDSK} file overwrites the old version of the \textit{EQDSK} file, in case the user wants to use the latest output \textit{EQDSK} file as as source for profiles or equilibrium geometry for a new iteration.} So, if the user wants to use or have an \textit{EQDSK} file of higher resolution their values need to be changed from the default, i.e. 60, and go higher in value, e.g. 513.\\~\\
\textit{CheasePy} code will call the \cheasepyinline{create\_namelist} function and pass the list of parameters which will in turn create the \textit{chease\_namelist} file required  by \textit{CHEASE} code.

\subsubsection{Modes for CheasePy Running}
There are several options for a user who runs \textit{CheasePy} that can be accessed by setting the value of \cheasepyinline{runmode} in \cheasepyinline{cheaseVals} dictionary variable:
\paragraph{Option 1:} To setup the environment for \textit{CHEASE} code to run by creating the required files, i.e. \textit{EXPEQ}, \textit{EXPTNZ}, and \textit{chease\_namelist}, and start send the right command to the system to run the \textit{chease\_hdf5} executable file in the terminal to establish the MHD equilibrium for the current input data. At the end of each successful run, \textit{CheasePy} code will generate several plots to compare the input and output profiles in the \textit{cheaseresults.pdf} file by calling \cheasepyinline{plot_chease} function.
\paragraph{Option 2:} In case of unsuccessful run that doesn't produce \textit{cheaseresults.pdf} file, the second option that \cheasepyinline{runmode} can take instruct \textit{CheasePy} code to search for the output HDF5 files from \textit{CHEASE} code and plot their contents against the input profiles.
\paragraph{Option 3:} This option instructs \textit{CheasePy} to search for all the available input and output files from any previous \textit{CHEASE} run and remove all of these files. Hence, the current folder will contains only \textit{chease\_hdf5} and \textit{runchease.py}. This option is helpful when a user wants to start a new \textit{CHEASE} run, because when \textit{CheasePy} starts running it searches first for any files from a previous run and automatically continue from where that previous run stopped.\\~\\
When a user selects \textbf{option 1} for the \cheasepyinline{runmode} in \cheasepyinline{cheaseVals} which instructs \textit{CheasePy} to run \textit{CHEASE} code, the user can setup the environment using \cheasepyinline{removeinputs} and \cheasepyinline{removeoutputs} options before the run starts, and the user have three other options:
\paragraph{Option 1:} If the user sets both \cheasepyinline{removeinputs} and \cheasepyinline{removeoutputs} options to \textit{yes}, \textit{CheasePy} will remove all the input and output files related any previous run before running \textit{CHEASE} code. This requires a path to the shot files to be provided by the user for \textit{CheasePy} to copy them into the current folder before starting a new run.
\paragraph{Option 2:} If the user sets \cheasepyinline{removeinputs} to \textit{no} and \cheasepyinline{removeoutputs} to \textit{yes}, the \textit{CheasePy} code will keep the input files and remove all the output files from any previous run, then \textit{CheasePy} code will start an new run using the available input files in the current directory without any need to copy these files from their source directory.
\paragraph{Options 3:} If the user sets both \cheasepyinline{removeinputs} and \cheasepyinline{removeoutputs} options to \textit{no}, \textit{CheasePy} will keep all the input and output files from any previous run and use the simulation data of the last iteration from that previous run as the initial condition of the current run and starts the simulation from there. This option is helpful when a previous run crashes or reaches the maximum wall time before the end of the simulation, so \textit{CheasePy} can restarted from where it left.\\~\\
The next option the user need to set in the \cheasepyinline{cheaseVals} dictionary is the path to the folder/directory of the shot measurements to copy these profile and equilibrium geometry files to the current directory. This step is required if the input and output files from a previous run were removed. The \textit{CheasePy} user needs to remember that the filename of profile and equilibrium geometry files needs to start with folder name followed by an `\textit{underscore}' then the type of the data files. For example, for a shot folder named \textit{DIIID\_174082}, the profiles and equilibrium geometry files should be named: \textit{DIIID\_174082\_EXPTNZ}, \textit{DIIID\_174082\_EXPEQ}, \textit{DIIID\_174082\_CHEASE}, \textit{DIIID\_174082\_EQDSK}, or \textit{DIIID\_174082\_PROFILES}.\\~\\
The last option the \textit{CheasePy} user has to set before calling the \cheasepyinline{cheasepy} function is the \cheasepyinline{cheasemode} entity in the \cheasepyinline{cheaseVals} dictionary which can be set to any of the following options:
\paragraph{Option 1:} Setting \cheasepyinline{cheasemode} to `\textbf{1}' tells \textit{CheasePy} to reconstruct the MHD equilibrium using \textit{CHEASE} code using the input profiles and equilibrium geometry. If the \cheasepyinline{iterTotal} entity in \cheasepyinline{srcVals} is set to a number `\textbf{n}', \textit{CHEASE} will rerun for `\textbf{n}' iterations, and the source files may change depending on the length of profile's sources list.
\paragraph{Option 2:} Setting \cheasepyinline{cheasemode} to `\textbf{2}' tells \textit{CheasePy} to reconstruct the MHD equilibrium from the input profiles and equilibrium geometry using \textit{CHEASE} code. Then, \textit{CheasePy} will use the error in estimating the total toroidal current to correct the parallel current ($I_{||}$ or $J_{||}$), and use this corrected quantity as input to the \textit{CHEASE} next iteration. \textit{CheasePy} will estimate that correction in every iteration until the error is small enough to stop the simulation.
\paragraph{Option 3:} Setting \cheasepyinline{cheasemode} to `\textbf{3}' tells \textit{CheasePy} to reconstruct the MHD equilibrium from the input profiles and equilibrium geometry using \textit{CHEASE} code. Then, \textit{CheasePy} will use the error in estimating the total toroidal current to correct the pressure component ($P$ or $P'$), and use this corrected quantity as input to the \textit{CHEASE} next iteration. \textit{CheasePy} will estimate that correction in every iteration until the error is small enough to stop the simulation.\\~\\
The \cheasepyinline{current\_correction} and \cheasepyinline{pressure\_correction} functions in the \textit{cheasepy.py} are called automatically every time a correction in \textit{parallel current} (option 2) or pressure (option 3), respectively, needs to be done. The \textit{CheasePy} user is free to do any modifications to these functions to handle the correction in a more suitable way for the case under study. But it is recommended that the user treat these functions as a template and do not change the variable name of the returned values for the \textit{CheasePy} to run properly and give relevant results.

\subsection{Plotting the Outputs}
When \textit{CheasePy} code completes the current run it plots some fields such as the density and temperature profiles, the current flux profile, the magnetic field, the toroidal current, etc., with all figures packed together in a single PDF file.\footnote{It is easy to add plots for other fields those are not plotted by copying the chunk of code used to plot another field and replace the name of the existing field with the new field name. You need also to rename the figure object and add it to the PDF file.} The user can plot the fields in the outputs of a previous run of \textit{CheasePy} directly without rerun the code by by calling the \cheasepyinline{plot\_chease(OSPATH,skipfigs)} function which takes the path to the HDF5 files as the first arguments (\cheasepyinline{OSPATH}), and the number of of skipped figures as the second argument (\cheasepyinline{skipfigs}). The number of skipped figures is useful when the \textit{CHEASE} does large number of iterations and the user need to plot only few of them, so the user can set \cheasepyinline{skipfigs} to a number of figures that will be skipped between the plotted ones. When passing a path to a folder as the first argument to the \cheasepyinline{plot\_chease} function, it searches for all files named \textit{chease\_iterxxx.h5}, parse them one by one, extract all the fields and coordinates, and then plot the selected fields. However, if you passed the path to a specific HDF5 file, it plots only the fields in this single file.
\section{Future Plans}
\textit{CHEASEPY} code is under continuous development to add new features to the code. However, this version of the code is stable and supposed to give no issues. In case the user has any concern or need to add more features to the code you may contact Ehab Hassan at ehab@utexas.edu. Otherwise, users can add these features themselves but they supposed to inform the original developers with these added features.

\section{Acknowledgment}
I definitely appreciate the huge contribution of Gabrielle Merlo with his valuable time through a continuous discussion, feedback, and guidance since in October 2018 to have this work done. Also, I would like to thank Olivier Sauter for replying to several questions via email, and responding to our concerns about understanding the workflow in \textit{CHEASE} code.\\~\\
I need also to thank Mike Kotschenreuther, David Hatch, and Yasutaro Nishimura for feedback and discussion about the best workflow for \textit{CheasePy}.\\~\\
Also, thanks should be given to Michael Halfmoon and Max Curie for running test cases and provide me with valuable feedback and comments that help me to put up this user guide together in this form.

\section{References}
\begin{itemize}
\item Lütjens, Hinrich, Anders Bondeson, and Olivier Sauter. "The CHEASE code for toroidal MHD equilibria." Computer physics communications 97.3 (1996): 219-260.
\item Sauter, O., and S. Yu Medvedev. "Tokamak coordinate conventions: COCOS." Computer Physics Communications 184.2 (2013): 293-302.
\end{itemize}

\end{document}